# Estimation of spin relaxation lengths in spin valves of In and In$_2$O$_3$ nanostructures

Keshab R Sapkota,[1,2] Parshu Gyawali,[2] Ian L. Pegg,[1,2] and John Philip[1,2]

[1]*Department of Physics, The Catholic University of America, Washington, DC 20064, USA*

[2]*The Vitreous State Laboratory, The Catholic University of America, Washington, DC 20064, USA*

We report the electrical injection and detection of spin polarized current in lateral ferromagnet-nonmagnet-ferromagnet spin valve devices, ferromagnet being cobalt and nonmagnet being indium (In) or indium oxide (In$_2$O$_3$) nanostructures. The In nanostructures were grown by depositing pure In on lithographically pre-patterned structures. In$_2$O$_3$ nanostructures were obtained by oxidation of In nanostructures. Spin valve devices were fabricated by depositing micro magnets over the nanostructures with connecting nonmagnetic electrodes via two steps of e-beam lithography. Clear spin switching behavior was observed in the both types of spin valve devices measured at 10 K. From the measured spin signal, the spin relaxation length ($\lambda_N$) of In and In$_2$O$_3$ nanostructures were estimated to be 449.6 nm and 788.6 nm respectively.

## I. INTRODUCTION

A lateral spin valve is a device structure in which ferromagnetic spin injector, nonmagnetic spin channel and ferromagnetic spin detector are fabricated in lateral geometry. Resistance of the device depends upon the spin states of the electrons passing from the spin injector to the detector, which can be utilized to understand spin relaxation phenomena in the spin channel.[1] Recently, considerable attention has been drawn in the search of spin valve devices based on various spin channel materials to investigate their possible spintronics applications. Use of In, a metal, and In$_2$O$_3$, a n-type transparent semiconductor, as a spin channel is particularly interesting as it will provide a comparative study of a metal based and its derivative semiconductor based spin transport as well as the latter is a potential candidate for transparent spintronics applications.[2] In and $In_2O_3$ have been extensively studied for their electrical and optical properties but spin based phenomena on these materials are little known.[3–5] Earlier, Blatt et al. reported magnetoresistance in thin In wires and Weiher et al. reported magnetoresistance in single crystalline $In_2O_3$.[4,6] Recently, Androulakis et al. studied the possibility of thin layer of $In_2O_3$ film as a tunnel barrier in the spintronics devices.[7] Besides the attempts of the magnetoresistance studies in In and $In_2O_3$, the realization of spin valves based on the nanostructures of these materials will broaden the understanding the spin transport on these materials as well as explore the possibilities of their spintronics applications.

In this article, we present the successful realization of In and $In_2O_3$ nanostructures based lateral spin valve devices. The In nanostructures were prepared by depositing In on pre- lithographically patterned structures in ultrahigh vacuum chamber. The $In_2O_3$ nanostructures were obtained by annealing In nanostructure in properly controlled $O_2$ environment. The spin valve devices were fabricated using multistep of electron-beam lithography. Spin injection and detection in the In or the $In_2O_3$ based spin valves were achieved by (Co) micromagnets. The low temperature measurements of spin valves exhibited successful spin injection and detection from which spin relaxation lengths of In and $In_2O_3$ were estimated.

## II. EXPERIMENTS

In order to prepare In nanostructures, patterns were developed on Si/SiO2 wafer by standard e beam lithography process. Pure indium (99.99%, obtained from Alfa Aesar) was deposited over the patterns inside ultra-high vacuum (UHV) chamber keeping pressure $10^{-8}$ torr. The scanning electron microscope (SEM) image of In nanostructure after liftoff is shown in Fig. 1(a). The grain formation on as deposited films was controlled by the wafer temperature, film thickness and deposition rate, and high quality conducting films were obtained. It was found that using optimal deposition rate and temperature of the wafer, continuous films were obtained for the thickness greater than 75 nm. Magnified SEM image of the In nanostructure, Fig. 1(b), shows the grainy but continuous deposited indium film. For the preparation of $In_2O_3$ nanostructures, oxidation of In nanostructures were carried out by annealing them at 350 $^0$C for 45 minutes in an atmosphere of 25 % of oxygen and 75 % of argon. The surface roughness before and after oxidation were found to be unchanged. The as prepared nanostructures of In and $In_2O_3$ were used to fabricate spin valves by multistep of electron beam lithography.

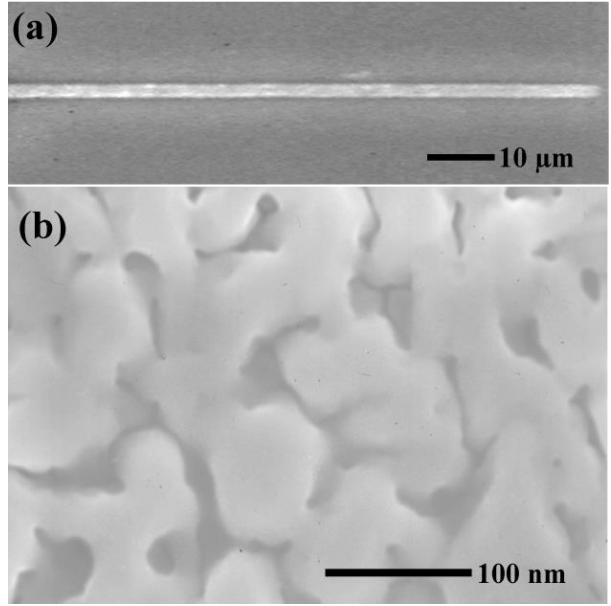

FIG.1. SEM image of (a) In nanostructure, (b) magnified view of the In nanostructure showing grainy surface of continuous film

For the characterization of $In_2O_3$ nanostructures, In films were deposited on glass substrates as reference samples along with the lithographically patterned wafer as discussed above. The oxidation of In film on reference samples was carried out along with the In nanostructures, cares were taken to maintain same oxidation conditions in both type of samples. The crystallographic and optical characterization of the $In_2O_3$ were carried out on the reference samples.

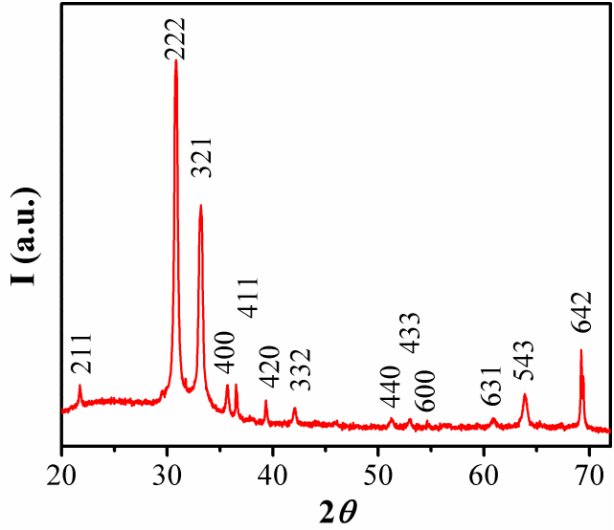 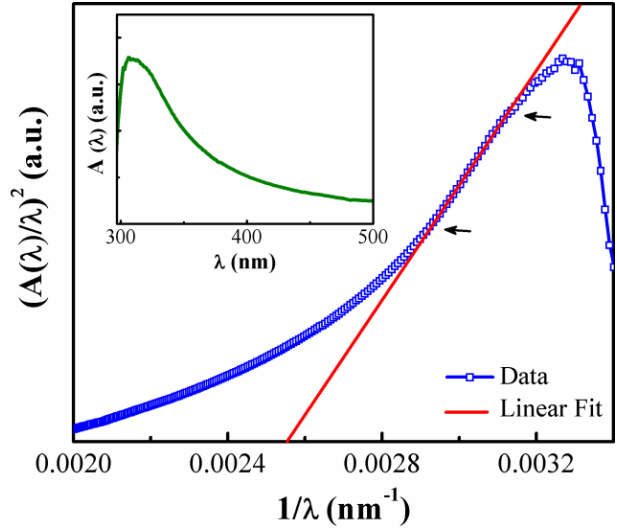

FIG. 2. XRD measurement of In$_2$O$_3$ film.   FIG. 3. Band gap estimation of In$_2$O$_3$. Inset shows the absorbance spectrum of In$_2$O$_3$.

## III.   RESULTS AND DISCUSSION

To study the crystal structure of In$_2$O$_3$, X-ray diffraction (XRD) analysis of the In$_2$O$_3$ reference samples was carried out. XRD spectrum, as shown in Fig. 2, revealed that In$_2$O$_3$ exhibited cubic structure of bixbyite Mn$_2$O$_3$ (I) type with a lattice parameter $a = 1.018$ nm which closely matches the reported lattice parameter for In$_2$O$_3$.[3]

Semiconducting property of In$_2$O$_3$ was investigated by spectrophotometry measurements of reference sample using the PerkinElmer spectrophotometer.  Measurements were done by scanning of In$_2$O$_3$ film with monochromatic wavelengths in the range of visible to UV region and the absorbance was recorded corresponding to scanned wavelengths. The obtained absorbance spectrum was then analyzed with the Tauc model to extract the value of bandgap for the In$_2$O$_3$. The absorption coefficient $\alpha(\lambda)$ for a crystalline semiconductor can be related to the incident photon energy $hc/\lambda$ near the band gap region by following equation:[8]

$$\alpha(\lambda)\frac{hc}{\lambda} = B\left(\frac{hc}{\lambda} - E_g\right)^m, \qquad (1)$$

where $hc/\lambda$, $B$ and $E_g$ incident photon energy, optical constant and optical bandgap respectively and $m$ is the index which can take different values of 1/2, 1/3, 2 and 3. The absorption coefficient $\alpha(\lambda)$ can be expressed in terms of absorbance $A(\lambda)$ according to Beer-Lambert's law as $\alpha(\lambda) = 2.303 \times A(\lambda)/d$, where $d$ is the thickness of the film.[9] Using Beer-Lambert law in equation (1), we can get,

$$\left(\frac{A(\lambda)}{\lambda}\right)^{1/m} = B_1\left(\frac{1}{\lambda} - \frac{1}{\lambda_g}\right), \qquad (2)$$

where, $B_1 = d \times B(hc)^{m-1}/2.303$, and $\lambda_g$ is wavelength corresponding to bandgap. Using equation (2), one can calculate bandgap by extrapolating linear region of the $(A(\lambda)/\lambda)^{1/m}$ versus $1/\lambda$ curve at $(A(\lambda)/\lambda)^{1/m} = 0$ and obtain $\lambda_g$. The best fitting is obtained for m = 2. Figure 3 shows the fitted data for the linear portion of $(A(\lambda)/\lambda)$ versus $1/\lambda$ curve with equation (2), the obtained value of bandgap from the fit is 3.1 eV. Inset of the Fig. 3 shows the as obtained absorbance data. In$_2$O$_3$ possess direct bandgap of about 3.6 eV and indirect bandgap of about 2.6 eV but the measured and calculated values have been reported to vary significantly.[3,10]

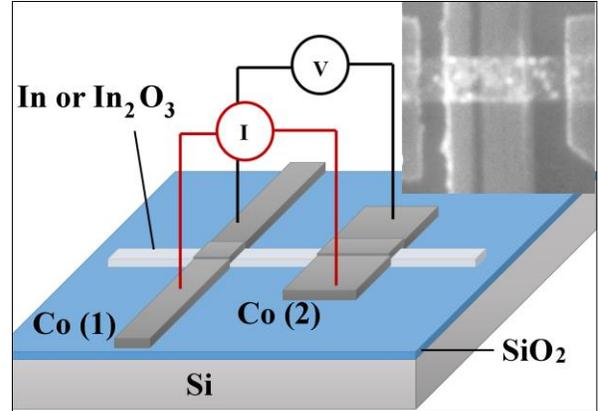

FIG. 4. Schematic and measurement scheme of spin valve. Inset shows the SEM image of the In or In$_2$O$_3$ nanostructure spin valve.

The spin valve devices of In and In$_2$O$_3$ nanostructures were fabricated by two steps of electron beam lithography. In the first step, two copper (Cu) electrodes were fabricated until near to the nanostructure where micromagnets connecting the nanostructure would make contacts. The outer side of Cu electrodes were available for wire-bonding. After completing Cu electrodes deposition and liftoff process, second step of electron beam lithography was carried to fabricate two micromagnetic electrodes of cobalt (Co) over the nanostructure connecting to Cu electrodes.

The two cobalt electrodes, labelled as Co(1) and Co(2), had the typical geometry of 1µm × 40µm and 3µm × 10µm respectively. Different aspect ratios for Co(1) and Co(2) were maintained to get different coercivities as required for the successful injection and detection of spin signals. In a typical device, the In or $In_2O_3$ nanostructure was 2 µm wide and 75 nm thick and the center to center length of the nanostructure between two Co electrodes was 2.5 µm. The spin valve measurement was carried out at 10 K by a standard ac lock-in technique with the excitation in the range of 1 µA to 20 µA, frequency was kept at 37 Hz. During the measurement, a magnetic field was applied parallel to the Co electrodes length axis and the field was scanned from negative to positive (forward scan) as well as from negative to positive (reverse scan) values. The typical field scanning range was between -2000 Oe to +2000 Oe. The measured spin signals from the In and $In_2O_3$ nanostructure spin valves are shown in Fig. 5(a) and 5(b) respectively. The observed spin valve effect in the devices can be explained in classic way. The current through a cobalt electrode, say Co(1), to spin channel (In or $In_2O_3$ nanostructure) injects spin polarized current in it. The spin polarization decays along the length of spin channel due to various spin scattering phenomena. The other cobalt electrode Co(2) detects the spin polarization on the current flowing from spin channel to it by exhibiting high or low resistance depending upon the magnetization direction on it. The state of high (low) resistance on device appears when magnetization directions of Co(1) and Co(2) are antiparallel (parallel). The state of high resistance appears as a crest on the plot of Fig. 5. The condition of parallel and antiparallel magnetization between Co(1) and Co(2) were achieved by utilizing the different magnetic coercivities of Co(1) and Co(2) introduced by magnetic shape anisotropy.[11]

Following the Fert and Lee model, the spin signal measurement of a spin valve in local configuration can be expressed as:[1,12]

$$\Delta R_l = 2\alpha^2 \frac{\lambda_N R_N}{L} exp\left(-\frac{L}{\lambda_N}\right), \qquad (3)$$

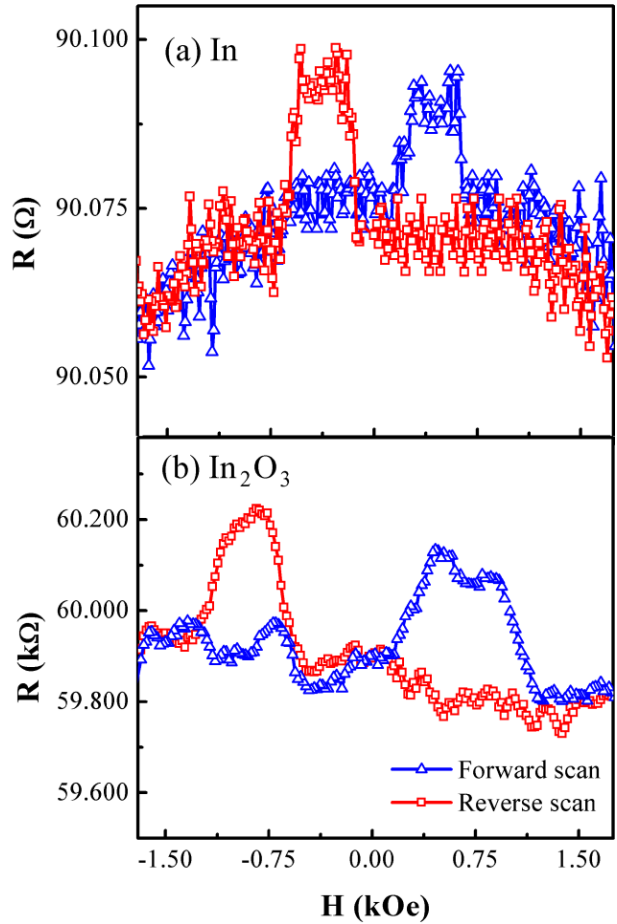

where, $\Delta R_l$ is spin signal (difference of resistance between antiparallel and parallel magnetization configuration, i.e. height of the crest from base line in Fig. 5), L is the length of the nanostructure (spin channel) between Co electrodes, $R_N$ is the resistance of the spin channel, $\lambda_N$ is the spin relaxation length, α is the injected spin polarization in the spin channel from Co electrode. Equation (3) shows the exponential decay of spin polarization along the spin channel length and represents the spin transport in $\lambda_N < L$ regime in which spin flip can occur before reaching to detector. Very low magnetoresistance, 0.024 % in case of In spin channel and 0.46 % in case of $In_2O_3$ spin channel from Fig. 5, supports the fact of spins flip and decay of spin polarized current before reaching to Co detector in our devices.[13–15]

FIG. 5. Spin valve measurement of In and $In_2O_3$ nanostructure devices.

Equation (3) can be rewritten as:

$$\frac{2\alpha^2 R_N}{\Delta R_l} = \frac{L}{\lambda_N} exp\left(\frac{L}{\lambda_N}\right). \quad (4)$$

Equation (4) takes the form of "Lambert W" function ($z = W(z)\ e^{W(z)}$ where $W(z)$ is Lambert W function). Solution of equation (4) gives spin relaxation length $\lambda_N$ in terms of Lambert -W function.

$$\lambda_N = \frac{L}{W(2\alpha^2 R_N/\Delta R_l)} \quad (5)$$

The Table I shows the calculated values of $\lambda_N$ using experimental data in equation (5). The value of $R_N$ is taken as the average value of the section of base line curve under spin signal of the magnetoresistance curve. Spin polarization $\alpha$ injected to the spin channel (In or $In_2O_3$) depends on the spin polarization of Co electrode which is taken to be 0.42 as reported in ref. 16.[16] Length (L) of the In or $In_2O_3$ nanostructure, which is center to center distance between Co electrodes, is measured as 2.5 µm.

TABLE I. Calculation of spin relaxation length ($\lambda_N$) for In and $In_2O_3$.

| Nanostructure | $\Delta R_l$ (Ω) | $R_N$ (Ω) | W function | $\lambda_N$ (nm) |
|---|---|---|---|---|
| In | 0.022 | 90.074 | 5.56 | 449.6 |
| $In_2O_3$ | 279 | 59928 | 3.17 | 788.6 |

It is worth noting that, different authors report different values of spin polarization in Co that in most cases vary between 0.35 to 0.42, giving $\lambda_N$ between 476.0 nm to 449.6 nm for In and 862.4 nm to 788..6 nm for $In_2O_3$.[16–19] Further, the injected spin polarization from Co electrode to spin channel can be diminished by the ferromagnetic- nonmagnetic junction where conductivity mismatch can play a role.[20] In these scenarios we would have smaller $\alpha$ giving larger value of $\lambda_N$ in the calculation. Nevertheless, the calculations in this experiment ascertains $\lambda_N$ not less than the tabulated values for In and $In_2O_3$ nanostructures, which is an important information for the design of spin based devices for the successful spin signal detection.

## IV. CONCLUSIONS

In conclusion, lateral spin valve devices of In and $In_2O_3$ nanostructures were successfully fabricated. The spin valve devices exhibited successful spin injection and detection at 10K. Spin signals were measured at local geometry and spin relaxation length of the In and $In_2O_3$ nanostructures were estimated. The $In_2O_3$ nanostructure shows longer spin relaxation length than In nanostructure, suggesting $In_2O_3$ is superior for spintronic device applications.

## ACKNOWLEDGEMENTS

This work was supported by National Science Foundation under ECCS-0845501 and NSF-MRI, DMR-0922997. We thank Cathy Paul for carefully reading the manuscript.